# A framework for the measurement and prediction of an individual scientist's performance

Endel Põder

Institute of Psychology, University of Tartu, Estonia

E-mail: endel.poder@ut.ee

## Abstract
Quantitative bibliometric indicators are widely used to evaluate the performance of scientists. However, traditional indicators do not much rely on the analysis of the processes intended to measure and the practical goals of the measurement. In this study, I propose a simple framework to measure and predict an individual researcher's scientific performance that attempts to take into account the main regularities of publication and citation processes and the requirements of practical tasks. Statistical properties of the new indicator – a scientist's personal impact rate – are illustrated by its application to a sample of Estonian researchers.

## Introduction

Quantitative measurement of scientific output has started to play an important role in the lives of scientists. However, traditional bibliometric indicators are not well adapted to the practical tasks for which they are used.

Using simple counts of publications and citations to measure productivity and research quality of an individual scientist assumes that each scientific paper has a single author. This was once a reasonable idea, but it is fundamentally misleading today (Lindsey, 1980; Price, 1981; Põder, 2010). Perhaps everybody agrees that it is not correct to equate the contribution of a single author of an article with the contribution of any of the 1000 co-authors of a similar article. Although it is well known how to calculate unbiased measures in that case, the majority of users simply ignore the problem of multiple authorship.

The popular h-index (Hirsch, 2005) is based on an amusing mathematical idea of combining publication and citation counts, which is, however, arbitrary and unsupported by any theory or data (Lehmann et al, 2006, Van Eck & Waltman, 2008). There is no rational argument why publishing, for example, 10 papers that receive 10 citations each should be valued higher than publishing five papers that receive 20 citations each. This indicator ignores the problem of multiple authors as well (Schreiber, 2008). A large number of "improved" or alternative indicators have been proposed (Panaretos & Malesios; 2009, Bornmann et al, 2011). However, there is no consensus on which of these indicators are really useful, how to select a correct one, or how to combine them, for practical tasks. Frequently, the experts of bibliometrics recommend to "take into account" different contextual factors not included in the indicators themselves (Panaretos & Malesios, 2009; Bornmann & Marx, 2014). One may conclude that the measurement of scientific performance is necessarily subjective.

It is important to determine the goal of our measurement. When selecting candidates for an academic position or making decisions about financial support, the main objective is to predict future performance. We have to estimate a researcher's ability to produce new qualitative papers within some future period. This is fundamentally different from the prediction of one's cumulative citation score or h-index (Hirsch, 2007; Acuna et al, 2012), which are determined primarily by past performance. Obviously, the cumulative indicators are not good for the practically relevant prediction. Also, it is impossible to compare researchers of different ages using these indicators. A more practical strategy is using some fixed intervals and time series format that allows analyzing the past and predicting future performance (Fiala, 2014; Schreiber, 2015).

In this study, I propose a framework for the measurement and prediction of an individual scientist's performance. It is based on a simple analysis of the behavioural processes that produce bibliometric data and of the requirements of usual practical tasks. It is intended to avoid some obvious problems of traditional bibliometric measures.

## Theory

I suppose that there are two elementary processes that can be modelled more or less separately: producing publications and collecting citations. These processes can be related to different behavioural characteristics of individual researchers.

Scientific articles can be produced either individually or in larger or smaller research groups. A larger group produces a larger number of articles per unit of time, assuming that productivity of the members of different groups is the same. The straightforward measure of an individual's productivity is the number of articles published per unit of time divided by the number of people in the group. If we have reliable information on the proportions of individual contributions of the co-authors, we can use these data for a more accurate estimation of each individual's productivity. Usually, researchers participate and publish different articles in different groups and we have to sum an individual's contributions over all these groups.

We can estimate a scientist's personal productivity for an interval with duration of $d$ (e.g., for last five years):

$$p = \frac{\sum_{i=1}^{n} \frac{1}{a_i}}{d} \text{ (articles per year)}, \qquad (1)$$

where $n$ is number of articles published (co-authored) by a given researcher during the interval $d$, and $a_i$ is the number of authors of publication $i$.

It is reasonable to suppose that an individual's productivity is relatively stable but may change as a result of experience, ageing, and working conditions.

The primary unit that collects citations is an article. Individual citations of an article are more or less independent and can be viewed as a Poisson process, but different articles have very different rates of citation (Stewart, 1994). On average, the citation rate of an article reaches its maximum within two to three years after publication and then declines gradually afterwards (Fig. 1). I follow the usual assumption that the rate of citation is a proxy for the quality of the article (although it is likely corrupted by different kinds of irrelevant variability). The average citation rate does not change dramatically during several years, and we can use different available intervals to estimate citation rate of an article. However, shorter intervals are more affected by Poisson variability.

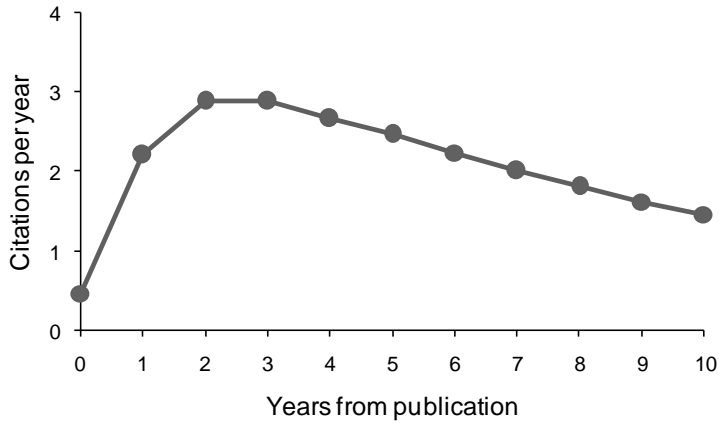

**Figure 1. Average time course of citation rate** (based on 23177 articles published by Estonian researchers, 1981-2012)

Undoubtedly, some people tend to publish articles of higher quality than others. Assume that every scientist can be characterized by a respective parameter (aptitude or talent for scientific research). With only single-author articles, we could measure the aptitude of a scientist by the mean (or median) citation rate of his/her articles (e.g., citations per article per year).

With co-authors, the average citation rate of the authored articles could be used as an approximation of the individual's aptitude as well. However, note that articles with different numbers of co-authors do not carry an equal amount of information in this regard.

Obviously, citations of single-author papers reflect the quality of a single author's work better than citations of papers published with many co-authors. The effects of the other authors can be regarded as additional measurement noise that increases with the number of co-authors. Therefore, the weighted average of citation rates, with weights inversely proportional to the number of co-authors, seems to be a better option.

Thus, we can estimate the quality of the researcher's production within an interval of interest as the weighted average citation rate of the articles published during this interval

$$q = \frac{n \sum_{i=1}^{n} \frac{c_i}{a_i}}{\sum_{i=1}^{n} d_i \cdot \sum_{i=1}^{n} \frac{1}{a_i}} \quad \text{(citations per article per year)}, \qquad (2)$$

$d_i$ is the duration when publication $i$ is available for citation within the interval of interest (for example, if an article was published in the third year of the five-year interval of interest, we may assume that $d_i$ is two years), and $c_i$ is the number of citations the publication $i$ received during the interval $d_i$. Note that this formula also appropriately takes into account the reliability of the estimated citation rate based on articles with different durations $d_i$, using more samples from those with longer intervals.

This measure (aptitude to produce highly cited articles) may change during one's career, and estimates based on more recent publications are expected to be more useful for the prediction. On the other hand, we should not use too short intervals that include only a few publications and result in extremely high sampling errors.

The final measure of a scientist's performance is the product of productivity and average quality. This could be named the (current or predicted) personal impact rate of a researcher.

$$f = p \cdot q = \frac{n \sum_{i=1}^{n} \frac{c_i}{a_i}}{d \sum_{i=1}^{n} d_i} \quad \text{(citations per year squared).} \tag{3}$$

If we assume an invariant rate of publication, then the expected $\sum_{i=1}^{n} d_i = \frac{n \cdot d}{2}$ and the equation can be simplified further:

$$f = \frac{2 \sum_{i=1}^{n} \frac{c_i}{a_i}}{d^2}, \tag{4}$$

or (assuming a delay of 0.5 year from publishing to first citations)

$$f = \frac{2 \sum_{i=1}^{n} \frac{c_i}{a_i}}{d(d-1)}. \tag{5}$$

For practical purposes, these versions (equations 3-5) are almost identical, having correlations of 0.98 or higher, with each other.

The unit of measurement (citations per year squared) may look somewhat strange at first glance. However, it reveals an interesting parallel with Newtonian mechanics. The scientific impact (similar to a mechanical one) can be related to acceleration. Once you publish scientific papers, they collect citations without any further effort, such as objects moving from inertia. To accelerate this process, you have to make an impact – i.e., to publish new citable papers. The present indicator estimates your propensity for that.

It is interesting that the proposed two-process theory leads to a very simple formula for the personal impact rate that basically needs only the summation of fractional citation counts of the articles published during the period of interest (equations 4, 5).

Unlike cumulative measures such as total citation count or h-index, which can only increase, the proposed measure of impact rate may decrease as well. Additionally, this measure has an attractive additive property: we can calculate the impact of a research group, an institution, or a country simply by summing up the impacts of the individuals. It is easy to see that the quality component (weighted average citation rate) resembles the journal impact factor.

# Application to empirical data

I applied the reported theoretical ideas to a sample of Estonian scientists (N=1356). (According to aggregate indicators, Estonian science is not far from the world average (Allik, 2013)). In this section, statistical properties of the proposed measures are analysed and their ability to predict future performance is estimated.

## Methods

The sample of researchers was derived from a database of 23 177 articles co-authored by Estonian scientists (provided by Thomson Reuters, 2013, for Estonian Research Council). From the database of articles, the authors with Estonian addresses were selected. Because only the address of the first author could be unambiguously identified from the given data, the selected authors appeared (at least twice) as a first author of an article. It would be problematic to apply the present quantitative analysis to subjects with only a few occasional publications. Therefore, a threshold of having published at least five articles was imposed.

My goal was to design a sample of individual scientists with their full history of publications and citations. To designate all the publications of one person to a unique name, I tried to correct all different spellings of the person's name. To avoid pooling the publications from several people with identical names and initials, the frequent combinations of names and initials (which appeared more than once in the list of Estonian Research Information System) were excluded. Additionally, the names not found in the list of the Research Information System were excluded from the sample. The final size of the sample was 1356. In the present analysis, the publication and citation data from years 1983 to 2012 were used. An observation window of at least a few years is needed in order to estimate publication and citation rates with some minimum acceptable reliability. In the present analysis, the whole 30-year interval was divided into six five-year periods. For these periods, the proposed measures of personal performance were calculated, their statistical properties were described, and the predictability of future performance from the measures of preceding periods was analysed.

## Results and discussion

Similar to traditional bibliometric indicators, the proposed measures of personal productivity, estimated research quality, and impact rate have highly skewed probability distributions on a linear scale. For example, the mean personal impact rate for the last period (2008-2012) was 0.9, its standard deviation 1.8, and maximum 35.

However, these data can be well approximated by lognormal distributions, as depicted in Figure 2.

The standard deviations on the logarithmic scale are very similar (approximately 0.4 in decimal logarithms) for the productivity and quality measures, and remain remarkably invariant over the studied interval of 30 years.

The productivity and quality measures were nearly independent (very small negative correlations, with one out of 6 reaching statistical significance, $p<0.01$). This appears to support the assumption about two separable traits that characterise individual researchers.

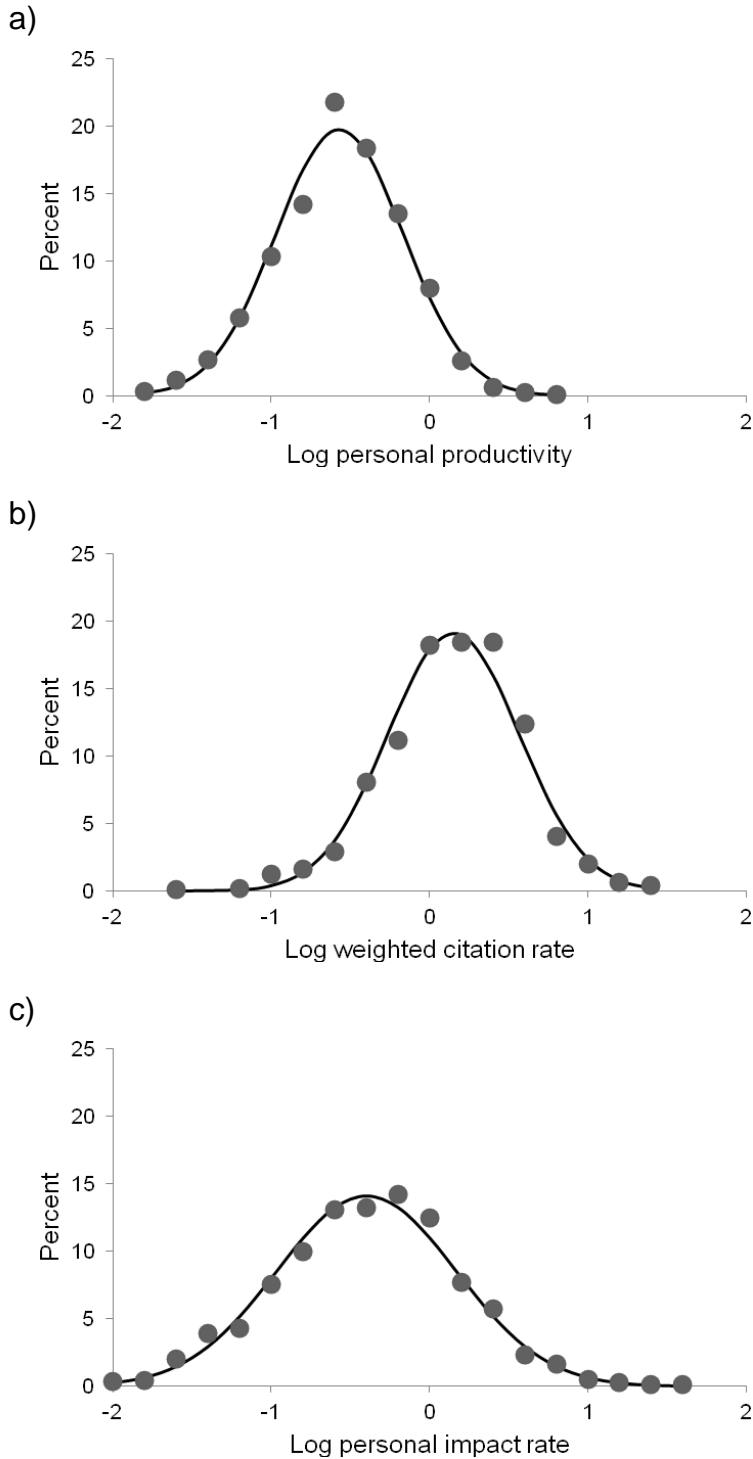

**Figure 2. Frequency distributions of the proposed indicators.** Log personal productivity (a), log weighted citation rate (b), and log personal impact rate (c) for the sample of Estonian researchers for the period of 2008-2012. Lines are best-fitted normal distributions.

Naturally, the personal impact rate, which is the product of these two factors, has larger standard deviation on the logarithmic scale (approximately 0.55).

I also calculated the correlations of the personal impact rate with two traditional indicators – count of publications, and the total number of citations received by these publications during the 5-year periods used in this study. The mean correlations between log transformed indicators were 0.72 and 0.85, for publication and citation counts, respectively. Although these correlations may look high, they don't imply that the correlated indicators measure the same quantity. Note that r=0.85 still leaves 27% of variance unexplained and allows considerable differences between the indicators. The correlations with usual cumulative measures were much lower – 0.60, 0.54, and 0.59, for counts of citations, publications, and h-index, respectively.

The predictive power of a bibliometric indicator can be estimated by the correlation of its current values with the values in some previous time period. All the measures proposed in this study, – personal productivity, quality, and impact rate, – have moderate correlations (approximately 0.4 to 0.5) with their own values in the preceding five-year period, across all the periods considered. The productivity measure seems to be slightly more predictable (mean correlation 0.52) compared with the quality measure (mean correlation 0.42). The mean correlation for the personal impact rate was 0.45 ($R^2$=0.20). Figure 3a depicts a scatter-plot example for the impact rate measure in two consecutive time periods. Figure 3b shows how predictability decreases with an increasing interval between the time periods.

Earlier, Hirsch (2007) and Mazloumian (2012) have attempted to predict bibliometric indicators that were similarly based on articles published within fixed observation periods. They used several traditional measures that were not adapted for the measurement of personal performance and they had different intervals of prediction. Overall, the correlations reported in their studies are comparable with the present results. The most notable difference was that predictability of a quality measure (citations per article) was much worse compared with other indicators reported in Hirsch (2007). In the present study, only a small difference in the predictability of quality and productivity measures was found.

However, I would not agree with the pessimistic interpretation of the reported moderate correlations in these studies (Mazloumian, 2012). Although $R^2 = 0.2$ leaves the larger part of variability in future performance unpredictable, the people selected from the higher end of the scale in the preceding period definitely have more chances of being successful in the future period (see Figure 3a). It is rational to use this information in decision making.

Finally, it is important to realize that predictive power, although a desired property of indicators, does not tell anything about the validity of an indicator. Note that a wrong indicator can be as good or a better predictor of its own future values compared to a correct indicator, and it may also predict better the future values of other (wrong) indicators. The validity of the indicators proposed in this study relies upon the transparent theoretical assumptions and logical implementation.

a)

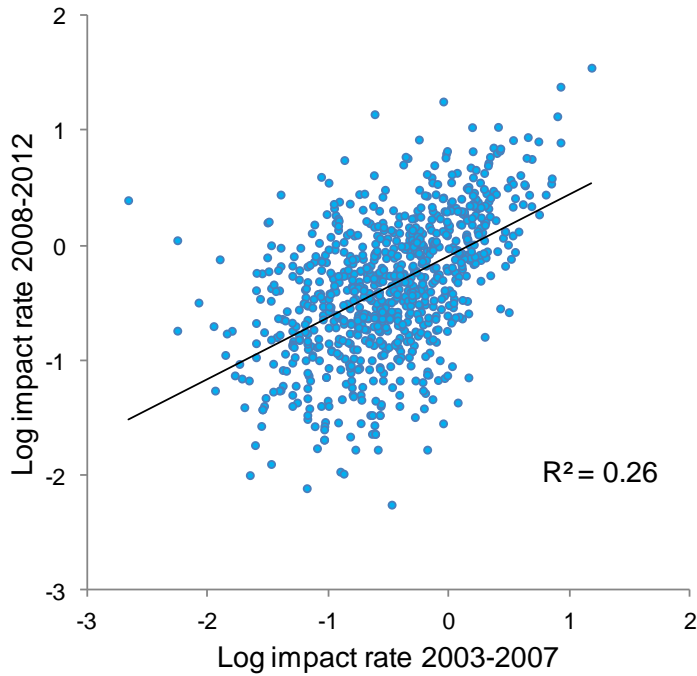

b)

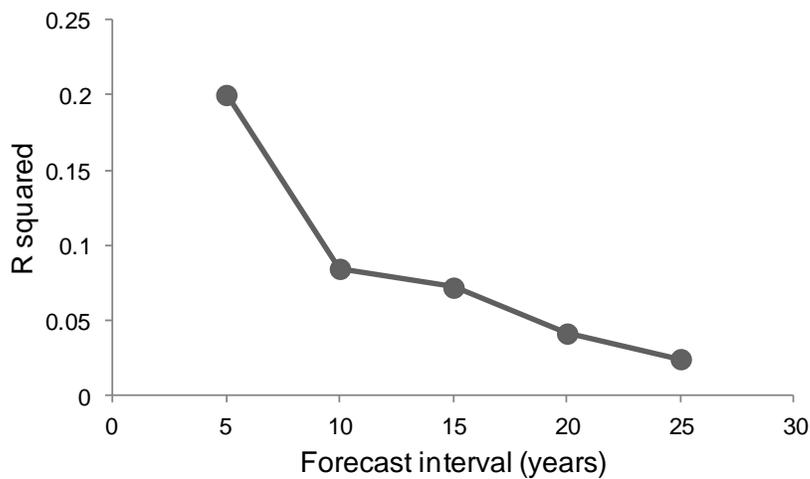

**Figure 3. Predictability of personal impact rate.** Relationship of personal impact rates in two consecutive five-year periods (a). Dots correspond to individual researchers, line indicates linear regression. Predictability of personal impact rate for a forthcoming period based on the measurement in an earlier period, as dependent on the interval between the periods (b).

# General discussion

This study proposed a simple theoretical framework for the analysis of publication and citation processes and their relationship to supposed characteristics of individual researchers. Based on these theoretical ideas, a new indicator of an individual scientist's performance was developed.

To adequately analyse publications with multiple authors, the present framework uses indicators based on fractional counts of publications and citations (Lindsey, 1980; Price, 1981). While dividing publications between the team members to calculate individual productivity seems very natural, dividing citation counts looks more problematic. Citation count of an article is a measure of quality, and quality should not depend on the number of co-authors.
Consistent with this idea, the quality measure proposed in this study uses a formula of weighted averaging of full citation counts that does not affect the overall level of the indicator. However, the final measure of performance, which combines both quality and quantity, reduces to the summation of fractional citation counts.

The proposed measure of impact rate is similar to the citation indicator used in Essential Science Indicators by the Thomson Corporation. The main differences are the following: using fractional citation counts, normalisation by squared duration, and inclusion of all the publications of a scientist within an interval of interest, in the proposed indicator.

The proposed framework of two processes indicates that there are two sources of possible errors when attempting to measure a scientist's performance.
In recent years, it has become a standard recommendation that for a meaningful comparison across research fields and publication years, citation counts should be normalized by some average or expected values (Schubert & Braun, 1996; Radicchi et al, 2008; Waltman & Van Eck, 2013). This is a reasonable strategy for many practical tasks, and it is not difficult to use some kind of normalized citation counts in the present model. However, there is another source of irrelevant variability that is independent of citation statistics but related to differences in authorship patterns. It has been shown that normalisation by number of co-authors may reduce the variability of citation counts across researchers of different fields as well (Batista et al, 2006). The fractional indicators proposed in this study should have a similar effect. An example of the measure that uses both types of corrections is "fractional scientific strength" proposed by Abramo and colleagues (Abramo et al, 2013; Abramo & D'Angelo, 2014).

The popularity of the h-index is explained by its simplicity and insensitivity to a few "accidentally" successful publications. However, this indicator combines publication and citation counts in a rather arbitrary way and has no natural relationship to time series data of publication and citation. I believe that similar robustness can be achieved more naturally with traditional tools of mathematical statistics. It is just necessary to take into account the stochastic nature of the processes, short observation intervals and small sample sizes, use log-transformed scales and the median instead of the arithmetic mean where appropriate, and calculate confidence limits of the measurements and predictions.

The main motivation of this study was to present a theoretically grounded way to measure an individual scientist's performance. The inevitable uncertainty of future performance and different opinions about the criteria of success may create an impression that it is not really important regarding which bibliometric measures to use; none of them can accurately predict the success of an individual researcher. I think this is a bad idea. Massive use of a biased measure may have strong detrimental effects, regardless of its correlation with the valid parameter (Põder, 2010).

## Conclusions

This study proposed a measure of an individual scientist's performance that is based on plausible theoretical assumptions, takes into account the main regularities of publication and citation processes, and meets practical goals of the measurement of scientific performance.

The new measure avoids the most obvious problems of traditional bibliometric indicators when applied to the measurement of personal performance. It can be easily calculated from available data. Its predictive power is similar to that of traditional measures.

The proposed theoretical framework allows for further improvements from the use of more elaborate statistical methods and more detailed knowledge about publication and citation processes.